\documentclass[aps,prl,twocolumn]{revtex4-2}
\usepackage{amsmath,bm}
\usepackage{graphicx,epsfig,braket,amssymb,amsfonts,eufrak,tabularx,multirow,amsmath,graphicx,color}
\usepackage{newtxtext}

\newcommand{\bd}{\begin{displaymath}}
\newcommand{\ed}{\end{displaymath}}
\renewcommand{\v}[1]{{\bf #1}}
\newcommand{\bpm}{\begin{pmatrix}}
\newcommand{\epm}{\end{pmatrix}}
\newcommand{\nn}{\nonumber \\}

\begin{document}

\title{Generation of Nonreciprocity of Gapless Spin Waves by Chirality Injection}

\author{Gyungchoon Go}
\affiliation{Department of Physics, Korea Advanced Institute of Science and Technology, Daejeon 34141, Korea}

\author{Seunghun Lee}
\affiliation{Department of Physics, Korea Advanced Institute of Science and Technology, Daejeon 34141, Korea}

\author{Se Kwon Kim}
\email{sekwonkim@kaist.ac.kr}
\affiliation{Department of Physics, Korea Advanced Institute of Science and Technology, Daejeon 34141, Korea}

\begin{abstract}
In chiral magnets with intrinsic inversion symmetry breaking, it has been known that two spin waves moving in opposite directions can propagate at different velocities, exhibiting a phenomenon called magnetochiral nonreciprocity which allows for realizations of certain spin logic devices such as a spin-wave diode. Here, we theoretically demonstrate that the spin-wave nonreciprocity can occur without intrinsic bulk chirality in certain magnets including easy-cone ferromagnets and easy-cone antiferromagnets. Specifically, we show that nonlocal injection of a spin current from proximate normal metals to easy-cone magnets engenders a non-equilibrium chiral spin texture, on top of which spin waves exhibit nonreciprocity proportional to the injected spin current. One notable feature of the discovered nonreciprocal spin waves is its gapless nature, which can lead to a large thermal rectification effect at sufficiently low temperatures. We envision that nonlocal electric injection of chirality into otherwise nonchiral magnets may serve as a versatile route to realize electrically controllable magnetochiral phenomena in a wide class of materials.
\end{abstract}

\maketitle

\emph{Introduction.}\textemdash
In certain materials with broken inversion symmetry, it has been known that direction-dependent propagation of particles can occur~\cite{Tokura2018}. In particular, chiral magnetic materials, in which both the inversion symmetry and the time-reversal symmetry are broken, are known to harbor such a nonreciprocal transport of spin waves~\cite{Cheong2018, Cheon2018}. Because spin waves carry energy and angular momentum without accompanying the Joule heating,
the nonreciprocal spin-wave phenomenon can be exploited to realize energy-efficient spin devices such as spin-wave logic gates~\cite{Jamali2013} and spin-wave diodes~\cite{Chumak2015}. The nonreciprocal propagation of spin waves is attributed to the asymmetric dispersion $\omega(k) \neq \omega(-k)$, which, for long wave length modes, can often be written as
\begin{align}\label{asydis}
\omega(k) = \omega_0(k)  + C k,
\end{align}
where $\omega_0(k)$ represents the symmetric component and the $k$-linear term is responsible for the spin-wave nonreciprocity. The latter has been interpreted as the spin-wave Doppler shift that can be induced by the spin-polarized current in magnetic metals~\cite{Vlaminck2008,Vlaminck2010,Sekiguchi2012, Kim2021}, the magnetostatic interactions~\cite{An2013}, the Dzyaloshinskii-Moriya interaction~\cite{Melcher1973, Kataoka1987, Moon2013, Di2015a, Di2015b, Seki2016, Kikuchi2016, Iguchi2015, Seki2020, Ogawa2021}, and the phonon and magnon drags~\cite{Chumak2010, Yu2021}. We note that the previously discussed nonreciprocal spin-waves possess a finite energy gap.

Recently, it has been recognized that it is possible to electrically inject the chirality into nonchiral magnetic materials with easy-plane or easy-cone anisotropy, whose order parameter is characterized by U(1) azimuthal angle which we denote by $\phi$, in the context of superfluid spin transport~\cite{KonigPRL2001, Sonin2010, Chen2014a, Kim2016a, Qaiumzadeh2017, Evers2020, Zarzuela2021}. Experimental schemes for superfluid spin transport have been proposed by using the nonlocal spin injection from proximate normal metals via the spin Hall effect~\cite{Takei2014, Takei2015}; its realizations have been reported in antiferromagnetic insulator Cr$_2$O$_3$~\cite{Yuan2018} and quantum-Hall graphene antiferromagnet~\cite{Stepanov2018}. In superfluid spin transport, the spin supercurrent is proportional to gradient of the U(1) parameter $\phi'$, which is manifested as a chiral spin texture~\cite{Sonin2010}. Long-wavelength spin waves in easy-cone magnets in equilibrium have the symmetric linear dispersion: $\omega(k) = c |k|$, which is gapless since it is the Goldstone mode associated with spontaneous breaking of the U(1) symmetry. The ability to maintain a chiral spin texture in easy-cone magnets by a nonlocal spin injection leads us to wonder if the injected chirality can induce nonreciprocity in otherwise reciprocal gapless spin-wave spectrum.

\begin{figure}[t]
\includegraphics[width=86mm]{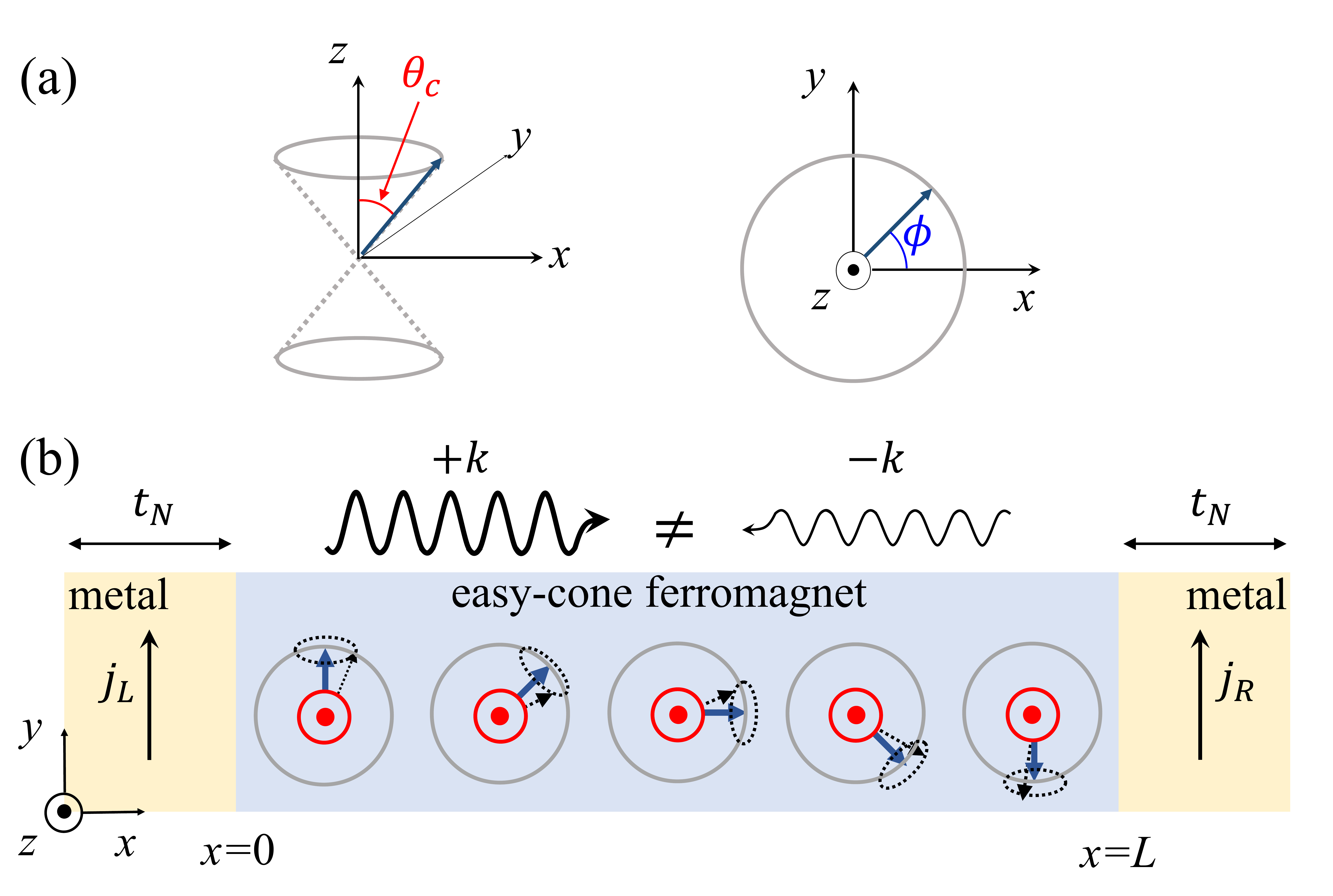}
\caption{(a) Schematic illustrations of ground states of easy-cone ferromagnets. The ground state manifolds are shown as solid gray lines, which form two cones with the cone angle $\theta_c$ from the $z$ axis. The azimuthal angle $\phi$ is arbitrarily chosen in a ground state, breaking the U(1) spin-rotational symmetry spontaneously. (b) Geometrical setup of the circuit configuration for the spin current injection. The spin configuration is shown for the symmetric case where the left and the right currents are identical ($j_L = j_R$). The chiral spin texture ($z$-component: red circled dot, $xy$-component: solid blue arrows) is induced by nonlocal spin injection from the two boundaries. The spin waves on top of the chiral spin texture (dotted black arrows) show a direction-dependent propagation characterized by the group-velocity difference between $+k$ and $-k$ (wavy arrows).}
\label{fig:1}
\end{figure}

In this Letter, we theoretically investigate spin waves in easy-cone magnets subjected to the nonlocal injection of the spin current. See Fig.~\ref{fig:1} for the illustration of the system. First, we show that the chiral spin texture induced by the spin-current injection leads to the asymmetric dispersion of gapless spin waves in an easy-cone ferromagnet. The induced spin-wave nonreciprocity is proportional to the gradient of the azimuthal angle $\phi'$ which corresponds to the spin supercurrent carried by the chiral spin texture~\cite{KonigPRL2001, Sonin2010, Chen2014a, Kim2016a, Qaiumzadeh2017, Evers2020, Zarzuela2021}. Therefore, the found nonreciprocity of spin waves can be interpreted as the effect of the spin-supercurrent-induced spin-wave Doppler shift, analogous to the Doppler shift of the Bogoliubov quasiparticle by the superfluid velocity in superfluids~\cite{Kohen2006}. We then expand our result to easy-cone antiferromagnets, which harbor two spin-waves modes, one gapless and the other gapful, unlike the ferromagnetic case possessing only a gapless mode. We show that the dispersions of both spin-wave modes become nonreciprocal in the presence of a chiral spin texture when a magnetic field is applied along the high-symmetry axis. For both ferromagnets and antiferromagnets, nonreciprocal spin waves are shown to give rive to nonreciprocal thermal transport within the Boltzmann transport theory. The obtained nonreciprocal spin waves are characterized by two features. First, the nonreciprocity of spin waves is induced nonlocally in nonchiral magnets by the injection of the spin current through the boundaries. Second, the obtained nonreciprocal spin waves are gapless, which can lead to stronger thermal rectification effect at sufficiently low temperatures compared to gapped counterparts. Our results show that the nonlocal induction of chirality via interfacial spin-current injection can engender magnetochiral phenomena in magnetic materials with no intrinsic chirality, which can be exploited to realize strongly nonreciprocal thermal transport under suitable conditions.

\emph{Spin-wave nonreciprocity in ferromagnets.}\textemdash
We begin by considering the following Hamiltonian density for a quasi-one dimensional uniaxial ferromagnet:
\begin{align}
\label{EqFM}
{\cal H} = \frac{A}{2} \left(\nabla{\v n}(x)\right)^2 - \frac{K_1}{2} n_z(x)^2 + \frac{K_2}{2} n_z(x)^4 \, ,
\end{align}
where $\v n(x) = {\v s(x)} /s$ is a unit vector along the local spin density $\mathbf{s}(x)$, $A$ is the exchange coefficient,
$K_1$ is the first-order effective anisotropy, and $K_2$ is the second-order anisotropy.
It is convenient to parametrize $\v n$ in the spherical coordinates: $\v n = (\sin\theta\cos\phi, \sin\theta \sin\phi, \cos\theta)$. Depending on the relative magnitude of $K_1$ and $K_2$, an equilibrium state is chosen among perpendicular magnetic anisotropy ($\mathbf{n}_0 = \pm \hat{\mathbf{z}}$), in-plane ($\mathbf{n}_0 \perp \hat{\mathbf{z}}$), and easy-cone (see Fig.~\ref{fig:1}(a)). In this work, we are interested in easy-cone phases, which are stabilized for $K_1 > 0$ and $K_2 > {K_1}/{2}$~\cite{Jang2019}.
Exemplary materials for easy-cone phases are Co/Pt multilayers, Ta/CoFeB/MgO, and NdFeB compounds~\cite{Stillrich2009, Shaw2015, Yamada1986, Papamantellos2003}.
The corresponding ground-state manifolds are two cones characterized by the polar angles (referred to as cone angles): $\theta_{c} = \left[\cos^{-1} (K_1 - K_2)/{K_2}\right]/2$ and $\pi - \theta_{c}$. The Hamiltonian [Eq.~\eqref{EqFM}] is invariant under the azimuthal-angle translation $\phi\rightarrow \phi + \Delta\phi$, possessing the U(1) spin-rotational symmetry~\cite{Kim2016a}. A ground state breaks the U(1) symmetry, which enables superfluid spin transport~\cite{Sonin2010}, as the spontaneous breaking of the U(1) phase symmetry allows for superfluid mass transport in superfluid helium.

Now, let us consider the setup shown in Fig.~\ref{fig:1}(b) where two heavy metals sandwich the easy-cone ferromagnet. The charge currents through the heavy metals inject a spin current into the magnet via the spin Hall effect. Throughout the Letter, we consider the situations where the two charge currents are the same ($j_L = j_R = j$) so that the realization of a static spin texture is allowed by the symmetry~\cite{Takei2015}. The dynamics is described by the Landau-Lifshitz-Gilbert (LLG) equation:
\begin{equation}
s \dot{\mathbf{n}} + \alpha s \mathbf{n} \times \dot{\mathbf{n}} = \mathbf{n} \times \left( A \mathbf{n}'' + K_1 n_z \hat{\mathbf{z}} - 2 K_2 n_z^3 \hat{\mathbf{z}} \right) \, ,
\end{equation}
where $s$ is the spin density, $\alpha$ is the Gilbert damping constant, and $'$ represents the derivative with respect to $x$. The boundary conditions are given by equating the spin current $J_z = - A \hat{\mathbf{z}} \cdot (\mathbf{n} \times \mathbf{n}')$ with the spin torque subtracted by the spin pumping:
\begin{equation}
- A \sin^2 \theta \phi' = \sin^2 \theta \left( j \vartheta \mp \gamma \dot{\phi} \right) \, ,
\end{equation}
where the upper and the lower sign correspond to the left ($x = 0$) and the right ($x = L$) interface, respectively~\cite{Tserkovnyak2014, Takei2015}. Here, $\vartheta$ is related to the effective interfacial spin Hall angle $\theta$ via $\vartheta\equiv \hbar \tan \theta / 2 e t_N$ ($e$: charge of an electron, $t_N$: normal metal thickness), $\gamma = \hbar g^{\uparrow \downarrow} / 4 \pi$, and $g^{\uparrow \downarrow}$ is the effective interfacial spin-mixing conductance. By solving the LLG equation in conjunction with the boundary conditions, one can show that the spin-current injection from the boundaries induces a static chiral spin texture characterized by the uniform spatial rotation of the azimuthal angle $\phi \rightarrow \phi_0 + \phi' x$ with $\phi' = - \vartheta j / A$, and a cone-angle deformation $\theta_{c} \rightarrow \left[\cos^{-1} (K_1 - K_2 + A (\phi')^2)/{K_2}\right]/2$~\footnote{See the Supplemental Material for the steady-state solution for chiral spin textures and the spin waves on top of it in various magnets, and the Holstein-Primakoff magnon descriptions of the spin-wave nonreciprocity.}. Here, note that the spin chirality $\phi' \propto j$ is induced nonlocally through the boundaries and it can be dynamically controlled by varying the charge current $j$ in the heavy metals.

\begin{figure}[t]
\includegraphics[width=86mm]{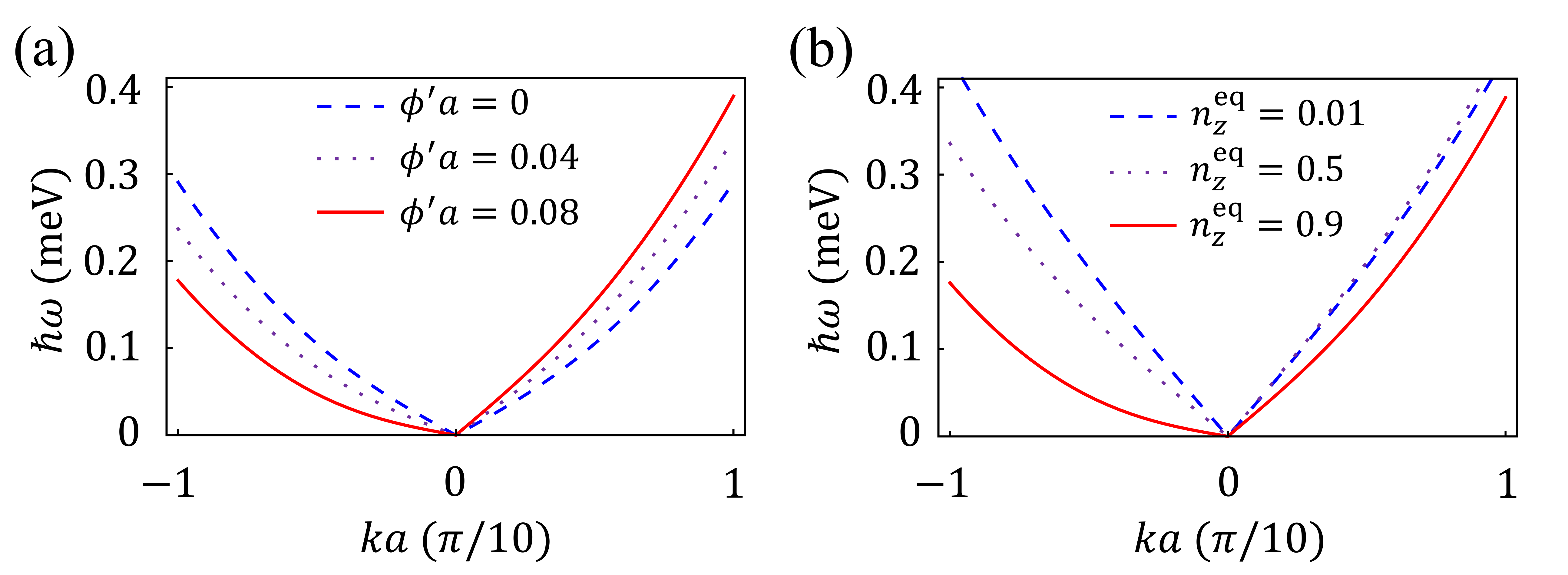}
\caption{Spin-wave dispersions [Eq.~(\ref{swdisp})] (a) for several spin chirality $\phi'$ with fixed $n_z^{\text{eq}} = 0.9$ and (b) for several values for $n_z^{\text{eq}}$ with fixed $\phi' a = 0.08$, where $a$ is the lattice constant. The asymmetry of the band between $k > 0$ and $k < 0$ increases as $\phi'$ and $n_z^\text{eq}$ increase.}
\label{fig:2}
\end{figure}

To investigate spin waves on top of the static spin texture, we divide $\v n(x, t)$ into the static profile $\v n_0(x)$ and the small fluctuations $\delta \v n(x, t)$, the latter of which can be written as $\delta \v n = n_\theta \hat{\boldsymbol\theta} + n_\phi \hat{\boldsymbol\phi}$,
where $\hat{\boldsymbol\theta} = [\partial_\theta \v n]|_{\mathbf{n}_0}$ and $\hat{\boldsymbol\phi} = \v n_0 \times \hat{\boldsymbol\theta}$.
By expanding the LLG equation to linear order in $n_\theta$ and $n_\phi$ with $\alpha = 0$, we obtain the equations of motion for spin waves:
\begin{equation}\label{eqndot}
\begin{aligned}
&s\dot n_\theta = -A n_\phi'' -  2A n_z^{\text{eq}} \phi' n_\theta',\\
&s\dot n_\phi = A n_\theta'' - K_{\phi'} n_\theta -  2A n_z^{\text{eq}} \phi' n_\phi',
\end{aligned}
\end{equation}
where $n_z^{\text{eq}} = \cos\theta_{c}$,  $K_{\phi'} = {{\bar K}_{1}(2 K_2 - {\bar K}_{1})}/{K_2}$, and ${\bar K}_{1} = K_1 + A (\phi')^2$ is a renormalized first-order anisotropy.
We note that the last terms in Eqs.~\eqref{eqndot} are proportional to the product of $n_z^{\text{eq}}$ and the spin-chirality parameter $\phi'$, which give rise to the spin-wave nonreciprocity as shown below. Inserting a plane-wave ansatz $n_\theta, n_\phi \propto e^{i( k x - \omega t) }$ into Eqs.~\eqref{eqndot} yields the following spin-wave dispersion:
\begin{align}\label{swdisp}
\omega(k) = \frac{1}{s}\sqrt{A k^2 \left[Ak^2 + K_{\phi'} \right]} + \frac{2 A}{s} n_z^{\text{eq}}\phi' k.
\end{align}
When $\phi' =0$, the first term of Eq.~\eqref{swdisp} describes a reciprocal spin-wave mode without the spin-current injection: $\omega \approx v_0 |k|$ where $v_0 = \sqrt{AK_0}/s$ for sufficiently small $k$. It is the second term that creates the difference between right-moving spin waves $(k>0)$ and left-moving spin waves $(k < 0)$, which propagate at the velocities $v_+ = v_0 + \delta v$ and $v_- = v_0 - \delta v$, respectively, with $\delta v =  (2A n_z^\text{eq} \phi')/s$ for sufficiently small $k$. Note that the velocity difference is proportional to the spin-chirality $\phi'$ which was originally injected nonlocally from proximate metals via spin-current injection. This is our first main result: the injection of the spin current into easy-cone ferromagnets gives rise to the nonreciprocity in otherwise reciprocal spin waves. One notable feature of the obtained nonreciprocal spin waves is its gapless nature, which makes it distinct from previously discussed nonreciprocal but gapful spin waves in chiral magnets~\cite{Moon2013, Di2015a, Di2015b,Kikuchi2016}.

To plot the spin-wave dispersion numerically, we adopt material parameters of yttrium iron garnet (YIG) films: $A = 3.7 \times 10^{-7}$ erg/cm~\cite{Klinger2015}, $a = 1.23$ ${\rm nm}$, $s/\hbar = 0.65\times 10^{22}$ ${\rm cm}^{-3}$~\cite{Li2019}, where $a$ and $s$ are lattice constant and spin angular momentum density, respectively.
By using spin Hall angle of $\beta$-W thin film ($\theta = 0.3$)~\cite{Pai2012}, we estimate $\phi' \approx 0.08/a$ for $j = 2.4\times 10^8$ A/cm$^2$.
For the effective magnetic anisotropy, we use $K_1 = 2.9\times 10^6$ erg/cm$^3$~\cite{Li2019, Lee2016, Fu2017, Lin2020}. The second-order anisotropy constant is chosen to be $K_2 > K_1/2$.
Fig.~\ref{fig:2} (a) and (b) show the spin-wave dispersion for different spin-chirality parameter $\phi'$ and out-of-plane spin component $n_z^{\text{eq}}$. The spin-wave nonreciprocity increases as $\phi'$ and $n_z^{\text{eq}}$ increases, albeit there is a upper limit on the induced chirality $\phi'$ for the stability of the spin texture~\cite{Sonin2010, Landau1941, Konig2014, Chen2014, Kim2016b}.

From the symmetry viewpoint, the spin-wave nonreciprocity requires breaking of both inversion and time-reversal symmetries~\cite{Tokura2018}, which, in our model, are broken by the spin chirality $\phi'$ and the spin density in the $z$-direction $n_z^\text{eq}$, respectively~\cite{Note1}. This is the reason that the nonreciprocal term in Eq.~(\ref{swdisp}) is proportional to $n_z^\text{eq} \phi'$. In the Supplemental Material~\cite{Note1}, we show that the nonreciprocal spin waves can also be realized in easy-plane ferromagnets in the presence of a magnetic field. In this work, we focus on easy-cone ferromagnets because of their spontaneous magnetization in $z$-direction which obviates the need for an external field. Also, to confirm our continuum theory for the spin-wave nonreciprocity by using the discrete spin model, we provide its Holstein-Primakoff magnon description in the Supplemental Material~\cite{Note1}.

\emph{Spin-wave nonreciprocity in antiferromagnets.}\textemdash
Motivated by our result on nonreciprocal spin waves in easy-cone ferromagnets, let us now investigate an analogous phenomenon in easy-cone antiferromagnets. We consider a quasi-one-dimensional uniaxial bipartite antiferromagnet which can be described by the following Lagrangian density:
\begin{align}
{\cal L} = &s {\v m}\cdot({\v n}\times \dot{\v n}) - \frac{A}{2} (\partial_i \v n)^2\nn
& - \frac{{\v m}^2}{2\chi} + \frac{K_1}{2} n_z^2 - \frac{K_2}{2} n_z^4 - \v b \cdot \v m,
\end{align}
where $\v n = (\mathbf{m}_1 - \mathbf{m}_2)/2$ and $\v m = (\mathbf{m}_1 + \mathbf{m}_2)/2$ are respectively the staggered (N\'{e}el) order parameter and the uniform component of the two sublattice spin directions $\mathbf{m}_1$ and $\mathbf{m}_2$~\cite{Hals2011, Takei2014}, $s$ is the saturated spin density , $A$ is the exchange coefficient, $\chi$ is the spin susceptibility, $K_1(>0)$ and $K_2(>{K_1}/{2})$ are the first- and second-order anisotropy coefficients, respectively, and $\v b$ represents the external magnetic field.
The equations of motion for $\v n$ and $\v m$ are obtained by minimizing the action with constraints $|\v n| = 1$ and $\v n\cdot \v m = 0$:
\begin{eqnarray}
s \dot{\v n} &=& \frac{1}{\chi} (\v m\times \v n) + \v b \times \v n \, ,\\
s \dot{\v m} &=& \v n \times \left[ A \nabla^2 \v n + (K_1 n_z - 2K_2 n_z^3 ) \hat {\v z} \right] + \v b \times \v m \, .
\end{eqnarray}
In terms of the spherical coordinate fields, we write $\v n = (\sin\theta \cos\phi, \sin\theta \sin\phi, \cos\theta)$
and $\v m = m_\theta \hat{\boldsymbol\theta} + m_\phi \hat{\boldsymbol\phi}$.
In a similar way to the ferromagnetic case, we read a quasi-equilibrium solution when the two charge currents in the adjacent heavy metals are the same ($j_L = j_R = j$)~\cite{Takei2015}:
\begin{equation}
\begin{aligned}
\quad \phi' = - \frac{\vartheta j}{A} \, , &\quad \theta_c = \frac12 \cos^{-1} \left(\frac{\bar {K_1 } - K_2 - b_0^2 \chi}{K_2}\right) \, , \\
m_\phi = 0 \, , &\quad  m_\theta = -\chi b_0 \sin\theta \, ,
\end{aligned}
\end{equation}
where $b_0$ is a magnitude of perpendicular magnetic field $(\v b = -b_0 \hat {\v z})$ and ${\bar K_{1}} = K_1 + A(\phi')^2$.
To obtain the spin-wave solution, we consider small deviations $\psi(x, t) = \left(\delta\theta, \xi_\phi, \delta\phi, \xi_\theta \right)$ from the equilibrium state defined by
\begin{equation}
\begin{aligned}
& \phi \rightarrow \phi' x + \frac{\delta\phi}{\sin\theta_c} \, , \quad \theta \rightarrow \theta_c + \delta \theta \, , \\
& m_\phi \rightarrow \xi_\phi \, , \quad m_\theta \rightarrow -\chi b_0 \sin\theta + \xi_\theta \, .
\end{aligned}
\end{equation}
Then, the linearized equations of motion for $\psi \propto \exp(i k x - i \omega t)$ yields the eigenvalue problem $(s\omega)\psi = \hat h \psi$ with
\begin{align}\label{AFMham}
{ \hat h} = i
\left(
  \begin{array}{cccc}
    0 & \chi^{-1} & 0 & 0 \\
    -A k^2 - \kappa_{\phi'} & 0 & - 2 i k A n_z^{\text{eq}} \phi' & - 2 b_0 n_z^{\text{eq}} \\
    -b_0 n_z^{\text{eq}} & 0 & 0 & -\chi^{-1} \\
    - 2 i k A n_z^{\text{eq}} \phi' & b_0 n_z^{\text{eq}} & A k^2 & 0 \\
  \end{array}
\right),
\end{align}
where $\kappa_{\phi'} = ({\bar K_{1}} - \chi b_0^2)\left[2 (K_2 + \chi b_0^2) - {\bar K_{1}}\right] / K_2$. When the staggered order parameter lies on the flim plane ($n^{\text{eq}}_z=0$), the effective Hamiltonian [Eq.~\eqref{AFMham}]
is divided by two block diagonal matrices for two basis functions $\psi_1 = (\delta\theta, \xi_\phi)$ and $\psi_2 = (\delta\phi, \xi_\theta)$, which have eigenvalues $\epsilon_1(k) =  \sqrt{[A k^2 + \kappa_{\phi'}]/\chi}$ and $\epsilon_2(k) =  \sqrt{A k^2/\chi}$, respectively~\cite{Takei2014}. The former and the latter describe the gapful spin wave and the gapless spin wave, respectively, of easy-plane antiferromagnets. By turning on $\phi'$ and $n^{\text{eq}}_z$, these basis functions hybridize and the eigenvalues are disturbed. However, these hybridization does not induce the spin-wave nonreciprocity in the absence of the magnetic field $b_0 = 0$. We find that the spin-wave nonreciprocity is generated when all of $\phi'$, $n^{\text{eq}}_z$, and $b_0$ are finite as discussed below.
\begin{figure}[t]
\includegraphics[width=86mm]{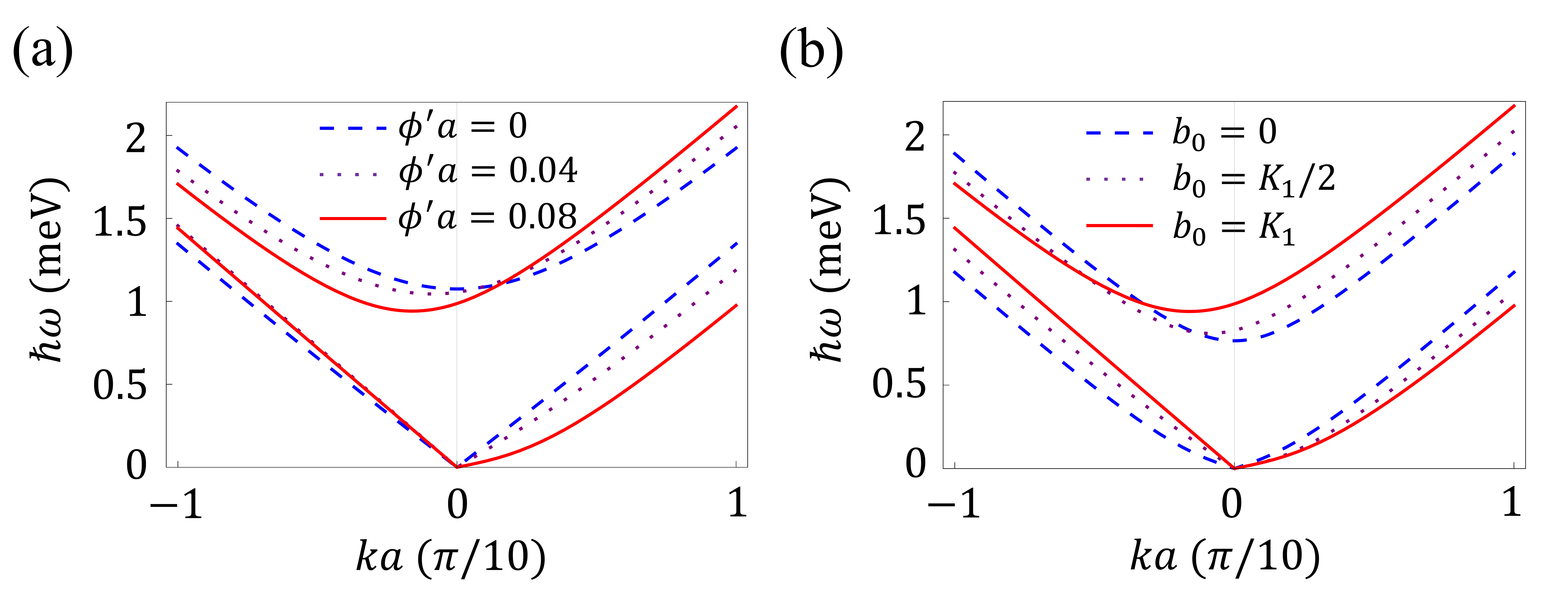}
\caption{Spin-wave dispersion of easy-cone antiferromagnets obtained from Eq.~(\ref{AFMham}) (a) for several spin chirality $\phi'$ and (b) for several magnetic fields $b_0$. Both gapless and gapful spin waves exhibit nonreciprocity. For (a), $K_2 = 1.71\times 10^6$ ${\rm erg/cm^3}$ and $b_0 = K_1$ are used. For (b), $K_2 = 1.71\times 10^6$ and $\phi' a = 0.1$ are used.}
\label{fig:3}
\end{figure}

To obtain the spin-wave dispersion numerically, we use $A = 3.7 \times 10^{-7}$ erg/cm, $a = 1.23$ ${\rm nm}$, $s/\hbar = 0.65\times 10^{22}$ ${\rm cm}^{-3}$,
$\theta = 0.3$, and $K_1 = 2.9\times 10^6$ ${\rm erg/cm^3}$. In Fig.~\ref{fig:3}(a) and (b), we show the spin-wave dispersion for different spin chirality $\phi'$ and the magnetic field $b_0$ when $n_z^{\text{eq}}$ is finite. When $\phi'$, $b_0$, and $n_z^{\text{eq}}$ are all finite, both upper and lower bands are asymmetric and the spin-wave nonreciprocity increases with $\phi'$ and $b_0$. This is our second main result: the nonlocal spin-current injection into an easy-cone antiferromagnet generates the nonreciprocity in both gapless and gapful spin waves when the magnet is subjected to an external field.

\begin{figure}[t]
\includegraphics[width=86mm]{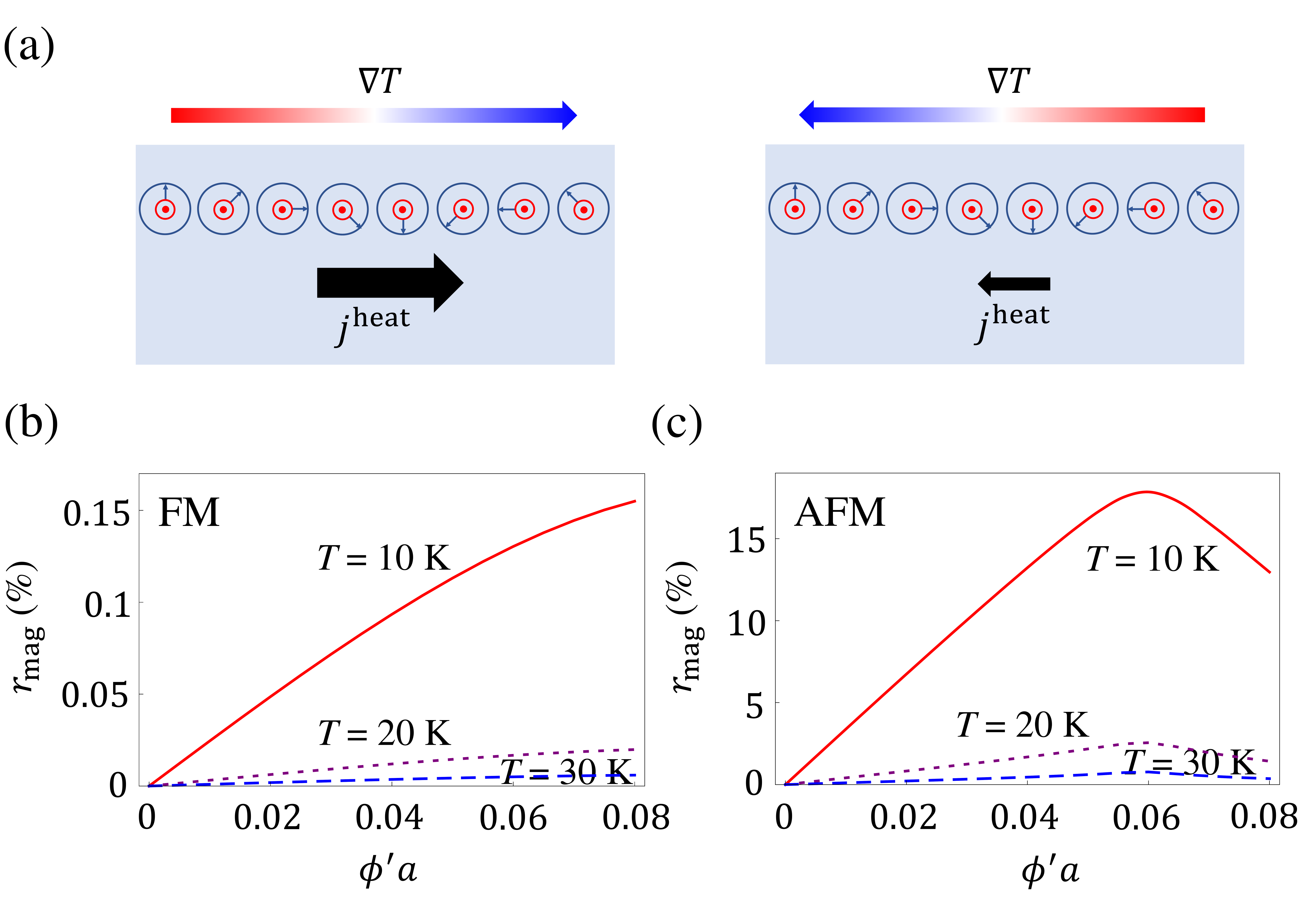}
\caption{(a) Schematic illustration of the nonreciprocal thermal transport.
Thermal rectification ratio of the magnons $r_\text{mag}$ [Eq.~(\ref{eq:rmag})] (b) in the ferromagnetic model (FM) and (c) in the antiferromagnetic model (AFM).
For (b), $K_2 = 1.9\times 10^6$ ${\rm erg/cm^3}$ is used. For (c), $K_2 = 1.71\times 10^6$ ${\rm erg/cm^3}$ and $b_0 = K_1$ are used.
}\label{fig:4}
\end{figure}

\emph{Thermal rectification.}\textemdash
The asymmetric spin-wave dispersion can give rise to the nonreciprocal heat transport, i.e., thermal rectification, as a nonlinear effect of the temperature gradient~\cite{Nakai2019, Hirokane2020}: $j^{\rm heat} = - \kappa_1 {\partial_x T} + \kappa_{2} {\partial^2_x T} + \tilde\kappa_{2} {(\partial_x T)^2}$, where $j^{\rm heat}$ is the heat current density, and $\kappa_1$ is the linear thermal conductivity. Here, $\kappa_{2}$ and $\tilde\kappa_{2}$ are the second-order thermal conductivities that result in the thermal rectification. Within the framework of the Boltzmann transport, the thermal conductivities are obtained for nonreciprocal spin waves in easy-cone ferromagnets and easy-cone antiferromagnets~\cite{Note1}. For a quantitative estimation, we use the following parameters: sample length $L = 1$ $\mu{\rm m}$, the temperature difference across the sample ${\delta T}/{T} = 10^{-2}$, and the spin-wave relaxation time $1/\tau = 10^{5}$ sec$^{-1}$~\cite{Spark1961}. Also, we assume the linear temperature gradient by setting $\partial_x^2 T = 0$. Figure~\ref{fig:4} shows the thermal rectification ratio for easy-cone ferromagnets and antiferromagnets defined by
\begin{align}
{r_{\rm mag} = \frac{j^{\rm heat} (\partial_x T >0) - j^{\rm heat} (\partial_x T <0)}{j^{\rm heat} (\partial_x T >0) + j^{\rm heat} (\partial_x T <0)}} \, .
\label{eq:rmag}
\end{align}
The thermal nonreciprocity $r_{\rm mag}$ increases with the current-induced spin chirality $\phi'$ as expected. We note that the rectification ratio in easy-cone antiferromagnets is much larger than that of easy-cone ferromagnets, which is due to the fact that the linear dispersion regime in the momentum space (where $\omega \propto k$), which gives the dominant contribution to the rectification, is much broader in antiferromagnets than in ferromagnets.

\emph{Discussion.}\textemdash
We have investigated spin waves in easy-cone magnets subjected to the nonlocal spin-current injection from the boundaries. We have shown that the injected spin current generates a chiral spin texture inside easy-cone magnets and spin waves on top of it exhibit finite nonreciprocity. Our results hold both for ferromagnets and for antiferromagnets. We would like to remark two features of the obtained nonreciprocal spin waves. First, the nonreciprocity of spin waves is generated by electrically injecting chirality into nonchiral magnets and thus allows for dynamical electric manipulation, which differs from previously discussed nonreciprocal spin waves in magnets with fixed intrinsic chirality. Second, due to the gapless nature of the nonreciprocal spin waves, the nonreciprocal thermal transport can be large at sufficiently low temperatures. Besides the discussed thermal transport, the nonreciprocal spin-wave transport can also be examined by other experimental methods such as the spin-wave spectroscopy~\cite{Vlaminck2008, Vlaminck2010, Seki2016} and the Brillouin light scattering~\cite{Di2015a, Di2015b}. We also would like to mention that the need to maintain the current in the heavy metals (in our setup shown in Fig.~\ref{fig:1}) to sustain the spin-wave nonreciprocity in our model can be obviated by replacing the heavy metals by metallic ferromagnets exhibiting a finite spin Hall effect~\cite{Miao2013, Tsukahara2014, Amin2019, Qu2020, Kim2020, Leiva2021} with easy-axis anisotropy along the $y$ axis, wherein the exchange field from the metallic magnet can fix the boundary magnetizations and thereby prevent the injected chiral texture from unwinding under favorable conditions. We hope that our result of nonlocal generation of the nonreciprocity of spin waves stimulates investigations of various novel magnetochiral phenomena that can be realized by nonlocal chirality manipulation of otherwise nonchiral materials.

\begin{acknowledgments}
We thank Kyung-Jin Lee and Shu Zhang for the useful discussions. G.G. acknowledges a support by the National Research Foundation of Korea (NRF-2019R1I1A1A01063594). S.L and S.K.K. were supported by Brain Pool Plus Program through the National Research Foundation of Korea funded by the Ministry of Science and ICT (NRF-2020H1D3A2A03099291), by the National Research Foundation of Korea(NRF) grant funded by the Korea government(MSIT) (NRF-2021R1C1C1006273), and by the National Research Foundation of Korea funded by the Korea Government via the SRC Center for Quantum Coherence in Condensed Matter (NRF-2016R1A5A1008184).
\end{acknowledgments}


\begin{thebibliography}{1}%
\makeatletter
\providecommand \@ifxundefined [1]{%
 \@ifx{#1\undefined}
}%
\providecommand \@ifnum [1]{%
 \ifnum #1\expandafter \@firstoftwo
 \else \expandafter \@secondoftwo
 \fi
}%
\providecommand \@ifx [1]{%
 \ifx #1\expandafter \@firstoftwo
 \else \expandafter \@secondoftwo
 \fi
}%
\providecommand \natexlab [1]{#1}%
\providecommand \enquote  [1]{``#1''}%
\providecommand \bibnamefont  [1]{#1}%
\providecommand \bibfnamefont [1]{#1}%
\providecommand \citenamefont [1]{#1}%
\providecommand \href@noop [0]{\@secondoftwo}%
\providecommand \href [0]{\begingroup \@sanitize@url \@href}%
\providecommand \@href[1]{\@@startlink{#1}\@@href}%
\providecommand \@@href[1]{\endgroup#1\@@endlink}%
\providecommand \@sanitize@url [0]{\catcode `\\12\catcode `\$12\catcode
  `\&12\catcode `\#12\catcode `\^12\catcode `\_12\catcode `\%12\relax}%
\providecommand \@@startlink[1]{}%
\providecommand \@@endlink[0]{}%
\providecommand \url  [0]{\begingroup\@sanitize@url \@url }%
\providecommand \@url [1]{\endgroup\@href {#1}{\urlprefix }}%
\providecommand \urlprefix  [0]{URL }%
\providecommand \Eprint [0]{\href }%
\providecommand \doibase [0]{https://doi.org/}%
\providecommand \selectlanguage [0]{\@gobble}%
\providecommand \bibinfo  [0]{\@secondoftwo}%
\providecommand \bibfield  [0]{\@secondoftwo}%
\providecommand \translation [1]{[#1]}%
\providecommand \BibitemOpen [0]{}%
\providecommand \bibitemStop [0]{}%
\providecommand \bibitemNoStop [0]{.\EOS\space}%
\providecommand \EOS [0]{\spacefactor3000\relax}%
\providecommand \BibitemShut  [1]{\csname bibitem#1\endcsname}%
\let\auto@bib@innerbib\@empty
%</preamble>
\bibitem [{Note1()}]{Note1}%
  \BibitemOpen
  \bibinfo {note} {See the Supplemental Material for the steady-state solution
  for chiral spin textures and the spin waves on top of it in various magnets,
  and the Holstein-Primakoff magnon descriptions of the spin-wave
  nonreciprocity.}\BibitemShut {Stop}%
\end{thebibliography}%


\begin{thebibliography}{99}


\bibitem{Tokura2018} Y. Tokura and N. Nagaosa, Nat. Commun. \textbf{9}, 3740 (2018).

\bibitem{Cheon2018} S. Cheon, H.-W. Lee, and S.-W. Cheong, Phys. Rev. B \textbf{98}, 184405 (2018).

\bibitem{Cheong2018} S.-W. Cheong, D. Talbayev, V. Kiryukhin, and A. Saxena, npj Quantum Mater. \textbf{3}, 19 (2018).

\bibitem{Jamali2013} M. Jamali, J. H. Kwon, S.-M. Seo, K.-J. Lee, and H. Yang, Sci. Rep. \textbf{3}, 3160 (2013).

\bibitem{Chumak2015} A. V. Chumak, V. I. Vasyuchka, A. A. Serga, and B. Hillebrands, Nat. Phys. \textbf{11}, 453 (2015).

\bibitem{Vlaminck2008} V. Vlaminck and M. Bailleul, Science \textbf{322}, 410-413 (2008).

\bibitem{Vlaminck2010} V. Vlaminck and M. Bailleul, Phys. Rev. B \textbf{81}, 014425 (2010).

\bibitem{Sekiguchi2012} K. Sekiguchi, K. Yamada, S.-M. Seo, K.-J. Lee, D. Chiba, K. Kobayashi, and T. Ono, Phys. Rev. Lett. \textbf{108}, 017203 (2012).

\bibitem{Kim2021} D.-H. Kim, S.-H. Oh, D.-K. Lee, S. K. Kim, and K.-J. Lee, Phys. Rev. B \textbf{103}, 014433 (2021).

\bibitem{An2013} T. An, V. I. Vasyuchka, K. Uchida, A. V. Chumak, K. Yamaguchi, K. Harii, J. Ohe, M. B. Jungfleisch, Y. Kajiwara, H. Adachi, B. Hillebrands, S. Maekawa, and E. Saitoh, Nat. Mater. \textbf{12}, 549–553 (2013).

\bibitem{Melcher1973} R. L. Melcher, Phys. Rev. Lett. \textbf{30}, 125 (1973).

\bibitem{Kataoka1987} M. Kataoka, J. Phys. Soc. Jpn. \textbf{56}, 3635–3647 (1987).

\bibitem{Moon2013} J.-H. Moon, S.-M. Seo, K.-J. Lee, K.-W. Kim, J. Ryu, H.-W. Lee, R. D. McMichael, and M. D. Stiles, Phys. Rev. B \textbf{88}, 184404 (2013).

\bibitem{Di2015a} K. Di, V. L. Zhang, H. S. Lim, S. C. Ng, and M. H. Kuok, J. Yu, J. Yoon, X. Qiu, and H. Yang, Phys. Rev. Lett. \textbf{114}, 047201 (2015).

\bibitem{Di2015b} K. Di, V. L. Zhang, H. S. Lim, S. C. Ng, M. H. Kuok, X. Qiu, and H. Yang, Appl. Phys. Lett. \textbf{106}, 052403 (2015).

\bibitem{Kikuchi2016} T. Kikuchi, T. Koretsune, R. Arita, and G. Tatara, Phys. Rev. Lett. \textbf{116}, 247201 (2016).

\bibitem{Iguchi2015} Y. Iguchi, S. Uemura, K. Ueno, and Y. Onose, Phys. Rev. B \textbf{92}, 184419 (2015). 

\bibitem{Seki2016} S. Seki, Y. Okamura, K. Kondou, K. Shibata, M. Kubota, R. Takagi, F. Kagawa, M. Kawasaki, G. Tatara, Y. Otani, and Y. Tokura, Phys. Rev. B \textbf{93}, 235131 (2016).

\bibitem{Seki2020} S. Seki, M. Garst, J. Waizner, R. Takagi, N. D. Khanh, Y. Okamura, K. Kondou, F. Kagawa, Y. Otani, and Y. Tokura, Nat. Commun. \textbf{11}, 256 (2020).

\bibitem{Ogawa2021} N. Ogawa, L. K\"{o}hler, M. Garst, S. Toyoda, S. Seki, and Y. Tokura, Proc. Natl. Acad. Sci. U.S.A. \textbf{118}, e2022927118 (2021).

\bibitem{Chumak2010} A. V. Chumak, P. Dhagat, A. Jander, A. A. Serga, and B. Hillebrands, Phys. Rev. B \textbf{81}, 140404(R) (2010).

\bibitem{Yu2021} T. Yu, C. Wang, M. A. Sentef, and G. E. W. Bauer, Phys. Rev. Lett. \textbf{126}, 137202 (2021).


\bibitem{Sonin2010} E. B. Sonin, Adv. Phys. \textbf{59}, 181 (2010).

\bibitem{KonigPRL2001} J. K\"onig, M. C. B\o{}nsager, and A. H. MacDonald, Phys. Rev. Lett. \textbf{87}, 187202 (2001).

\bibitem{Chen2014a} W. Chen and M. Sigrist, Phys. Rev. B \textbf{89}, 024511 (2014).

\bibitem{Qaiumzadeh2017} A. Qaiumzadeh, H. Skarsv{\aa}g, C. Holmqvist, and A. Brataas, Phys. Rev. Lett. \textbf{118}, 137201 (2017).

\bibitem{Evers2020} M. Evers and U. Nowak, Phys. Rev. B \textbf{101}, 184415 (2020).

\bibitem{Kim2016a} S. K. Kim and Y. Tserkovnyak, Phys. Rev. B \textbf{94}, 220404(R) (2016).

\bibitem{Zarzuela2021} R. Zarzuela, D. Hill, J. Sinova, and Y. Tserkovnyak, Phys. Rev. B \textbf{103}, 174424 (2021).

\bibitem{Takei2014} S. Takei, B. I. Halperin, A. Yacoby, and Y. Tserkovnyak, Phys. Rev. B \textbf{90}, 094408 (2014).

\bibitem{Takei2015} S. Takei and Y. Y. Tserkovnyak, Phys. Rev. Lett. \textbf{115}, 156604 (2015).

\bibitem{Yuan2018} W. Yuan, Q. Zhu, T. Su, Y. Yao, W. Xing, Y. Chen, Y. Ma, X. Lin, J. Shi, R. Shindou, X. C. Xie, W. Han, Sci. Adv. \textbf{4}, eaat1098 (2018).

\bibitem{Stepanov2018} P. Stepanov, S. Che, D. Shcherbakov, J. Yang, R. Chen, K. Thilahar, G. Voigt, M. W. Bockrath, D. Smirnov, K. Watanabe, T. Taniguchi, R. K. Lake, Y. Barlas, A. H. MacDonald, and C. N. Lau,
Nat. Phys. \textbf{14}, 907 (2018).

\bibitem{Kohen2006} A. Kohen, Th. Proslier, T. Cren, Y. Noat, W. Sacks, H. Berger, and D. Roditchev, Phys. Rev. Lett. \textbf{97}, 027001 (2006).

\bibitem{Jang2019} P.-H. Jang, S.-H. Oh, S. K. Kim, and K.-J. Lee, Phys. Rev. B \textbf{99}, 024424 (2019).

\bibitem{Stillrich2009} H. Stillrich, C. Menk, R. Fr\"{o}mter, and H. P. Oepen, J. Appl. Phys. \textbf{105}, 07C308 (2009).

\bibitem{Shaw2015} J. M. Shaw, H. T. Nembach, M. Weiler, T. J. Silva, M. Schoen, J. Z. Sun, and D. C. Worledge, IEEE Magn. Lett. \textbf{6}, 1 (2015).

\bibitem{Yamada1986} O. Yamada, H. Tokuhara, F. Ono, M. Sagawa, and Y. Matsuura, J. Magn. Magn. Mater. \textbf{54}, 585 (1986).

\bibitem{Papamantellos2003} P. S.-Papamantellos, C. Ritter, and K. H. J. Buschow, J. Magn. Magn. Mater. \textbf{260}, 156 (2003).

\bibitem{Tserkovnyak2014} Y. Tserkovnyak and S. A. Bender, Phys. Rev. B \textbf{90}, 014428 (2014).

\bibitem{Klinger2015} S. Klingler, A. V. Chumak, T. Mewes, B. Khodadadi, C. Mewes, C. Dubs, O. Surzhenko, B. Hillebrands, and A. Conca, J. Phys. D: Appl. Phys. \textbf{48}, 015001 (2015).

\bibitem{Li2019} G. Li, H. Bai, J. Su, Z. Z. Zhu, Y. Zhang, and J. W. Cai, APL Mater. \textbf{7}, 041104 (2019).

\bibitem{Pai2012} C.-F. Pai, L. Liu, Y. Li, H. W. Tseng, D. C. Ralph, and R. A. Buhrman, Appl. Phys. Lett. \textbf{101}, 122404 (2012).

\bibitem{Lee2016} S. Lee, S. Grudichak, J. Sklenar, C. C. Tsai, M. Jang, Q. Yang, H. Zhang, and J. B. Ketterson, J. Appl. Phys. \textbf{120}, 033905 (2016).

\bibitem{Fu2017} J. Fu, M. Hua, X. Wen, M. Xue, S. Ding, M. Wang, P. Yu, S. Liu, J. Han, C. Wang, H. Du, Y. Yang, and J. Yang, Appl. Phys. Lett. \textbf{110}, 202403 (2017).

\bibitem{Lin2020} Y. Lin, L. Jin, H. Zhang, Z. Zhong, Q. Yang, Y. Rao, and M. Li, J. Magn. Magn. Mater. \textbf{496}, 165886 (2020).


\bibitem{Landau1941} L. D. Landau, Zh. \'{E}ksp. Teor. Fiz \textbf{11}, 592 (1941).

\bibitem{Konig2014} J. K\"{o}nig, M. C. B{\o}nsager, and A. H. MacDonald, Phys. Rev. Lett. \textbf{87}, 187202 (2001).

\bibitem{Chen2014} H. Chen, A. D. Kent, A. H. MacDonald, and I. Sodemann, Phys. Rev. B \textbf{90}, 220401 (2014).


\bibitem{Kim2016b} S. K. Kim, S. Takei, and Y. Tserkovnyak, Phys. Rev. B \textbf{93}, 020402(R) (2016).



\bibitem{Hals2011} K. M. D. Hals, Y. Tserkovnyak, and A. Brataas, Phys. Rev. Lett. \textbf{106}, 107206 (2011).



\bibitem{Nakai2019} R. Nakai, and N. Nagaosa, Phys. Rev. B \textbf{99}, 115201 (2019).

\bibitem{Hirokane2020} Y. Hirokane, Y. Nii, H. Masuda, and Y. Onose, Sci. Adv. \textbf{6}, eabd3703 (2020).

\bibitem{Spark1961} M. Sparks, R. Loudon, and C. Kittel, Phys. Rev. \textbf{122}, 791 (1961).

%\bibitem{Sun2011} Q. Sun, Z. Jiang, Y. Yu, and X. C. Xie, Phys. Rev. B \textbf{84}, 214501 (2011).



\bibitem{Miao2013} B. F. Miao, S. Y. Huang, D. Qu, and C. L. Chien, Phys. Rev. Lett. \textbf{111}, 066602 (2013).

\bibitem{Tsukahara2014} A. Tsukahara, Y. Ando, Y. Kitamura, H. Emoto, E. Shikoh, M. P. Delmo, T. Shinjo, and M. Shiraishi, Phys. Rev. B \textbf{89}, 235317 (2014).

\bibitem{Qu2020} G. Qu, K. Nakamura, and M. Hayashi, Phys. Rev. B \textbf{102}, 144440 (2020).

\bibitem{Amin2019} V. P. Amin, J. Li, M. D. Stiles, and P. M. Haney, Phys. Rev. B \textbf{99}, 220405(R) (2019).

\bibitem{Kim2020} K.-W. Kim and K.-J. Lee, Phys. Rev. Lett. \textbf{125}, 207205 (2020).

\bibitem{Leiva2021} L. Leiva , S. Granville, Y. Zhang, S. Dushenko, E. Shigematsu, T. Shinjo, R. Ohshima, Y. Ando, and M. Shiraishi, Phys. Rev. B \textbf{103}, L041114 (2021).


\end{thebibliography}
\end{document}